\documentclass[aps,preprint,showpacs]{revtex4}
\usepackage{graphicx}

\begin{document}
\title{A New History-Driven Algorithm to Calculate the Density of States}
\author{Shijun Lei}
\affiliation{Purple Mountain Observatory, 2 West Beijing Rd, Nanjing Jiangsu, 210008 P.R. China}
\date{\today}

\begin{abstract}
We present a new Monte Carlo algorithm applying a history-driven mechanism for the calculation of the density
of states for classical statistical models. With the new method, detailed balance is naturally satisfied in limit and
the estimated density of state converges to the exact value. The new method could be easily evolved into the
multicanonical method to achieve high accuracy.
\end{abstract}
\maketitle
Among the various algorithms proposed to obtain the density of states ($g(E)$) through simulation \cite{Ferrenberg,Berg,Oliveira,
Lee,Wang02,Micheletti}, one effective strategy is to apply a history-driven mechanism to the Metropolis sampling Monte Carlo
(MC) simulation~\cite{Lee,Wang02,Micheletti}. The idea is to use the information earned from the history of the simulation (e.g.,
the histogram $H(E)$ accumulated) as a feedback to determine the transition probability of the future moves and hence the generation
of future states, so that the simulation is driven to sample the entire (or targeted) energy range efficiently and provides an estimation
of the density of states. In general, a MC trial move from energy $E_1$ to $E_2$ could be accepted according to the
Metropolis-Hastings transition probability:
\begin{equation}
p(E_2\rightarrow E_1)=\mathrm{min}\left[1,R(H)\right],
\end{equation}
where $R(H)$ is a certain function of the histogram and the core of the history-driven method. Starting with a random initial
guess, the history-driven simulation goes like:
\begin{equation}
\frac{H(E_1)}{H(E_2)}=\frac{g(E_1)p(E_2\rightarrow E_1)}{g(E_2)p(E_1\rightarrow E_2)}=\frac{g(E_1)}{g(E_2)}R(H),
\end{equation}
and the density of states that is $in~priori$ unknown could be estimated as:
\begin{equation}
\frac{g(E_1)}{g(E_2)}=\frac{H(E_1)}{H(E_2)}R(H)^{-1}.
\end{equation}
Certain correlation between the states generated in near sequence is necessary for a history-driven mechanism to take effect.
if, otherwise, the generation of a new state is completely independent to the history of the simulation, there wont be any
history-driven mechanism. 

If $R(H)$ is fixed during the simulation, Eq.~(2) describes the well known multicanonical simulation. A simple history-driven
mechanism is the multicanonical recursion~\cite{Berg96}. For example, the entropic sampling~\cite{Lee} adopts recursive periods
of multicanonical simulations:
\begin{equation}
\frac{H_{n}(E_1)}{H_{n}(E_2)}=\frac{g(E_1)}{g(E_2)}R_{n}(H).
\end{equation}
Starting with a random initial guess, such as $R_1(H)=1$, the transition probability is to be updated at the end of each period using
the histogram accumulated from that period:
\begin{equation}
R_{n+1}(H)=\frac{H_{n}(E_2)}{H_{n}(E_1)}R_{n}(H)=\displaystyle\prod_{i=1}^n\frac{H_i(E_2)}{H_i(E_1)}.
\end{equation}
To make the method practical, the histogram is initialized to be $H_i(E)=h_0\leq1$ at the beginning of each period \cite{Berg97}.

In practice, each period of the entropic simulation must be long enough so that a useful working estimation of the density of states
could be obtained for a final long run to achieve high accuracy. However, the long periods also slow the updating of $R_n(H)$
and reduce the efficiency of the simulation in sampling the targeted energy range, hindering the application of this method to a relatively
large system. A more timely updated transition probability could drive the simulation to go through the energy range faster. Obviously,
the limiting case of doing this is to update $R(H)$ right after every single MC move. With this single-move multicanonical recursion,
Eq.~(5) turns into: 
\begin{equation}
R(H)=\left(\frac{h_0+1}{h_0}\right)^{H(E_2)-H(E_1)}=f^{H(E_2)-H(E_1)},
\end{equation}
where $H(E)$ is the histogram maintained throughout the simulation and $f>1$ is a modification factor. (Note that $H_i(E)=h_0$ initially
and $H_i(E_1)=h_0+1$ if $E_1$ is visited after the only trail move of the $i^{th}$ period, so
$\frac{H_i(E_2)}{H_i(E_1)}=\left(\frac{h_0+1}{h_0}\right)^{-1}$. Similarly, $\frac{H_i(E_2)}{H_i(E_1)}=\frac{h_0+1}{h_0}$ if $E_2$ is visited.)
This is nothing but the transition probability used in the well-known Wang-Landau (WL) algorithm~\cite{Wang02}, which is actually a limiting
case of multicanonical recursion~\cite{Wangswendsen02}. With a properly chosen modification factor, usually $f=e=2.71828$, the WL
algorithm is very efficient in sampling throughout the energy range, conquering the probably dramatic energy landscape of the system. As noted
before, e.g., \cite{Belardinelli}, the estimation of density of states $\frac{g(E_1)}{g(E_2)}=\frac{H(E_1)}{H(E_2)} R(H)^{-1}=f^{H(E_1)-H(E_2)}$
(note that the WL algorithm results into a flat histogram) dose not converge to the exact value with a fixed $f$. High accuracy could also be
achieved for the WL algorithm by doing ``global updates" to $R(H)$ recursively (in addition to the ``local updates" that are taken after every
single move):
\begin{equation}
R_{n}(H)=R_{n-1}(H)f_{n}^{H_{n}(E_2)-H_{n}(E_1)}
\end{equation}
where $f_n$ is carefully reduced every time the histogram $H_n(E)$ is flat enough. To avoid saturation of errors, we use the $\ln(f)=1/t$
scheme~\cite{Belardinelli} for the WL algorithm in comparing with the new method in the rest of this paper.

Not confining to the multicanonical recursion, we may go back to Eq.~(2) and ask that whether other form of $R(H)$ could be found as
an effective history-driven mechanism to achieve high efficiency and accuracy. We find that $R(H)=\left(\frac{H(E_2)}{H(E_1)}\right)^{a}$
is a fairly good choice, where $a$ is properly chosen power index. Applying the new method, the simulation goes like:
\begin{equation}
\frac{H(E_1)}{H(E_2)}=\frac{g(E_1)}{g(E_2)}\left(\frac{H(E_2)}{H(E_1)}\right)^a.
\end{equation}
Detailed balance is naturally satisfied in limit and the histogram converges to:
\begin{equation}
\frac{H(E_1)}{H(E_2)}=\left(\frac{g(E_1)}{g(E_2)}\right)^{\frac{1}{a+1}}.
\end{equation}
In other words, the density of states could be estimated as:
\begin{equation}
\frac{g(E_1)}{g(E_2)}=\left(\frac{H(E_1)}{H(E_2)}\right)^{a+1}
\end{equation} which converges
to the exact value without any extra efforts of doing global updates.

According Eq.~(9), a sufficiently large power index $a$ should be used to make the histogram relatively flat and achieve efficiently sampling
throughout the targeted energy range. A simple choice of $a$ is to make $\exp(a)$ to be about the total number of configurations of the
system as that ensures a flatness of $\frac{H(E_1)}{H(E_2)}\lesssim e$. Tested for a $16\times16$ 2D Ising model with a total number of
configurations of $2^{256}\simeq e^{177}$, we find that any $a>200$ works almost as well in driving the simulation to sample throughout the
entire energy space using $\sim5.0\pm1.0\times10^5$ flips, while $a=100$ doubles the simulation effort with $\sim9.0\pm2.0\times10^5$
flips. As a comparison, the WL algorithm also takes $\sim5.0\pm1.0\times10^5$ flips with $ln(f)>0.1$ and $\sim8.0\pm2.0\times10^5$ flips
with $ln(f)=0.05$. It is worth to point out that, since the new method is as easily realized as the WL algorithm (see below), measurement in
MC trial moves (flips) is almost equivalent to that in CPU time in comparing these two methods. So the new method is as efficient as the WL
algorithm in sampling the energy space not only in MC trial moves but also in real CPU time.

Furthermore, we test the efficiency of the new method for a traveling salesman problem (TSP) consisting of 100 cities. For 100 points (``cities")
randomly distributed within a $1\times1$ square, the distances ($E$ in this problem) of the shortest and the longest cyclical path connecting
the 100 cities found after different simulation efforts are given in Tab.~\ref{tab1}. Trail moves proposed by \cite{Lin} are used and the path
length is binned into bins of width $0.01$ for the maintenance of $H(E)$ in the simulation. Fixed value of $f=e^{10}$ and $a=100000$ is
used for the WL and the new method, respectively. We repeat the tests for several different randomly generated distributions of the cities
and find similar results as shown in Tab.~\ref{tab1}. Again, the new method is as efficient as the WL method.

With the muticanonical method being a special case ($a=0$) of the new method, the latter shares many characters of the former. First of all,
the errors of the estimated density of states do not saturate even without doing global updates (i.e., with a fixed power index $a$). So the new
method could be useful for the estimation of the density of states in the cases where global updates are prohibited or not so feasible. For
example, both the WL simulation~\cite{Wust} and the new method simulation could be used to find out the energy space and the possible
energy entries of a system, which, in most cases, are not theoretically known as is the case of Ising model. For this purpose, the parameter
$f$ or $a$ shouldn't be reduced during the simulation, and the new method is superior to the WL method by providing an estimation of the
density of states that converges to the exact value. Moreover, the histogram resulted from the new method is not necessary to be flat. Actually,
with a proper guess of $g(E)$, the histogram could be roughly designed for various purposes.

\begin{figure}
\centering{\includegraphics[angle=270,width=1.0\textwidth]{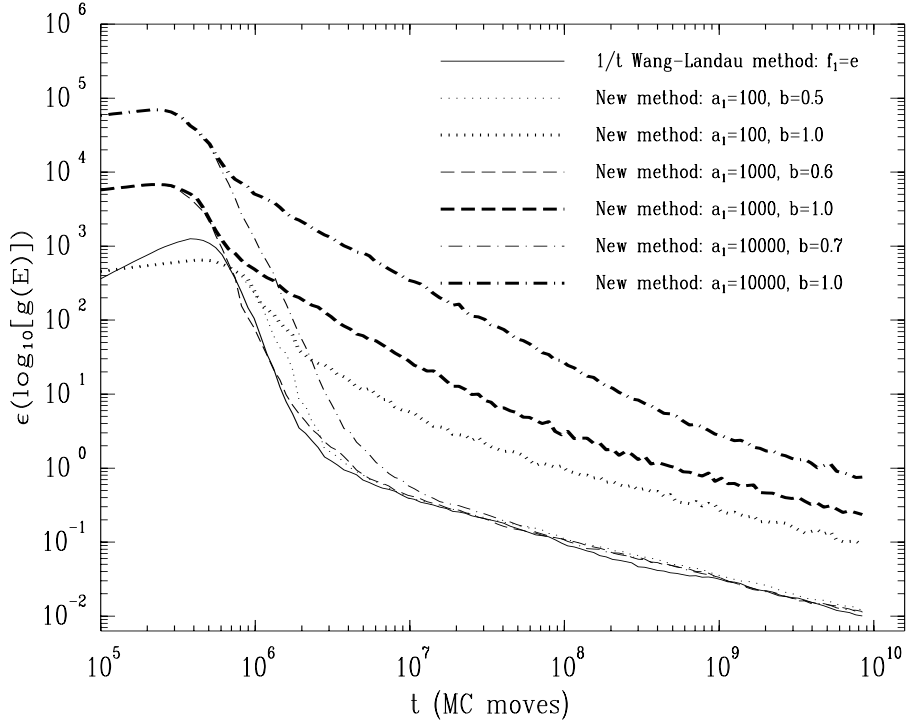}}
\centering\caption{Comparison of the accuracy of the new method and the $1/t$ WL method for $g(E)$ of the $16\times16$ 2D Ising
model. No sign of saturation of errors is found for the new method even without doing global updates ($b=1.0$). Applying the global
updates mentioned in the text, the new method is as accurate as the $1/t$ WL method. Each curve is averaged from 100 independent
runs. The error in the density of states is calculated as $\sigma[\lg g(E)]=\sqrt{\frac{\sum[\lg g(E)-\lg g_e(E)]^2}{n}}$, where $g_e(E)$
is the exact value and the summation is taken over all the energy entries of total number $n$. The normalization $g(0)=g_e(0)=\ln2$
is used for the calculation.}
\label{fig1}
\end{figure}

The new method is also closely related to the WL method. Given the well known result of the harmonic series that for $H(E)\rightarrow\infty$,
$\displaystyle\sum_{n=1}^{H(E)}\frac{1}{n}=\ln H(E)+c$, where $c$ is the Euler's constant , we have 
\begin{equation}
R(H)=\left(\frac{H(E_2)}{H(E_1)}\right)^a=e^{a[\ln H(E_1)-\ln H(E_2)]}\simeq
\exp\left(\displaystyle\sum_{n=1}^{H(E_1)}\frac{a}{n}-\displaystyle\sum_{n=1}^{H(E_1)}\frac{a}{n}\right)
\end{equation}
Comparing to Eq.~(6), we can see that only a small modification to the WL algorithm is needed to implement the new method, i.e., instead
of doing $\ln g(E)\leftarrow\ln g(E)+\ln f$, we do $\ln g(E)\leftarrow\ln g(E)+\frac{a}{H(E)}$ for the local updates. The histogram $H(E)$ must
also be maintained throughout the simulation, but with negligible calculating effort. Also, we see that the the power index $a$ here is an
equivalent of the modification factor $\ln f$ in WL method. Actually, for the 2D Ising model, we find that the fluctuation of the histogram is
proportional to $\frac{1}{\sqrt{a+1}}$, consistent with the finding in~\cite{Zhou03} that the fluctuation of the histogram resulted from WL
simulation is proportional to $\frac{1}{\sqrt{\ln f}}$. From Equ.~(10), we can predict that the error in the estimated density of states is proportional
to $\frac{a+1}{\sqrt{a+1}}=\sqrt{a+1}$, which is exactly what we see in Fig~\ref{fig1} in long run.

\begin{figure}
\centering{\includegraphics[angle=270,width=1.0\textwidth]{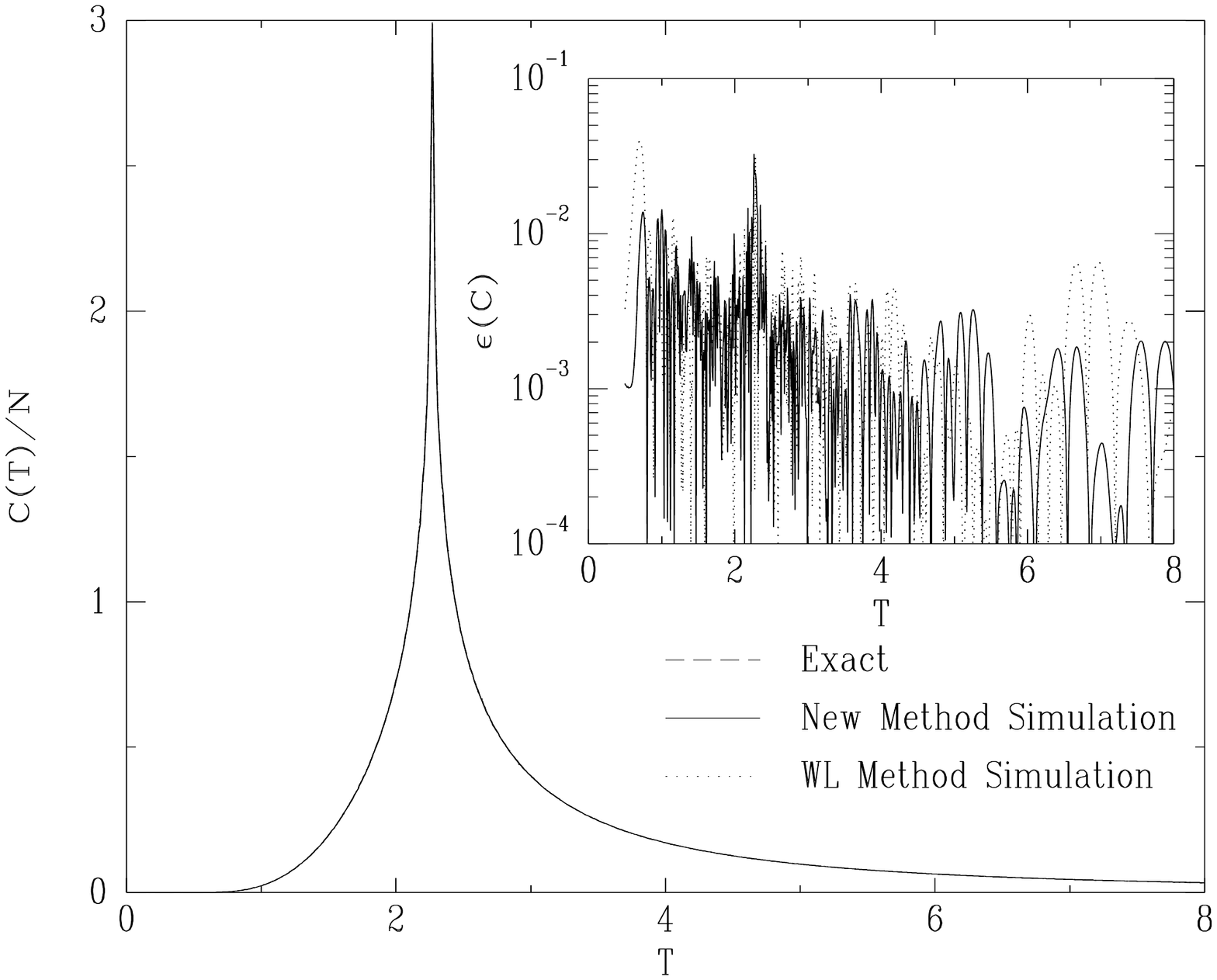}}
\centering\caption{Specific heat for the $256\times256$ 2D Ising model. The relative errors ($\epsilon(C)=|\frac{C-C_e}{C_e}|$, where
$C_e$ is the exact value) are shown in the inset. Both the new method simulation and the WL method simulation make use of a single
run of $5.12\times10^{12}$ MC moves.}
\label{fig2}
\end{figure}

If a better estimation of the density
of states is available, a smaller power index $a$ could be used to achieve high accuracy with reduced simulation effort. For example, using
the initial guess $g(E)=g(M)=C^{\frac{n}{2}+\frac{M}{2}}_n$ for the 2D Ising model (where $n$ is the total number of spins and $M$ the magnetization), $a$ could
an order of magnitude smaller than that needs for the initial guess $g(E)=1$. So we can reduce $a$ recursively when better estimation of the
density of states is available after a period of simulation and eventually turns the simulation into a multicanonical one. We here propose a
scheme of global updates of $R(H)$ that results into both high efficiency and accuracy. After the first period of simulation that samples the
targeted energy range, we update the transition probability
$R_1(H)=\left(\frac{H(E_2)}{H(E_1)}\right)^{a}$ globally into
\begin{equation}
R_2(H)=\left(\frac{H_1(E_2)}{H_1(E_1)}\right)^{a(1-b)}\left(\frac{H(E_2)}{H(E_1)}\right)^{ab},
\end{equation}
where $b\subset[0,1]$ is a modification factor. The histogram $H(E)$ is maintained throughout the simulation and whose value at the end of the
$n^{th}$ period is $H_n(E)$.  In general, after the entire energy space is sampled by a period of simulation, the following global update is taken,
\begin{equation}
R_{n+1}(H)=J_{n+1}(H)\left(\frac{H(E_2)}{H(E_1)}\right)^{ab^n},
\end{equation}
where
\begin{equation}
J_{n+1}(H)=J_n(H)\left(\frac{H_n(E_2)}{H_n(E_1)}\right)^{a(1-b)b^{n-1}}
\end{equation}
and $J_1(H)=1$.
For this purpose, a separated histogram $H_p(E)$ is maintained for each period and we apply a criterion $H_p(E)>H_{min}$ for ending
a period of simulation and doing global updates. Testing for the Ising model, we find that a modification factor of $b=0.5$ works robustly
with an $H_{min}\simeq l\times l$, where $l$ is the size of the Ising model. Seen from Eq.~(11), by doing the global updates, the new method
simulation seamlessly evolvesinto a muticanonical one as $b^n\rightarrow0$ and the second term on the right hand side vanishes when
$n$ goes large. Actually, we terminate the global update and explicitly turn the simulation into a muticanonical one when $ab^n<0.1$ in our
simulations.
'
We can see from
Fig.~\ref{fig1} that after applying the global updates, the new method is as accurate as the $1/t$ WL method. In Fig.~\ref{fig2}, we also show
the specific heat $C(t)$ calculated from the $g(E)$ for the $1024\times1024$ Ising model, which are obtained from simulations of the same
length but applying the new method or the WL method. Again, the new method is as accurate as the $1/t$ WL method. We also point out
that the global update scheme we proposed here is even more easily realized than the $1/t$ scheme in the WL method as no calculation
of $t$ is needed.

Comparing to the $1/t$ WL method, the only disadvantage we find for the new method is that it does not try to maintain a flat histogram
self-adaptively. The histogram may get not so flat and reduce the efficiency of simulation in long run. But we find this problem can be easily
solved by using proper values of $a$, $b$, $H_{min}$. The guide line is to use a larger $b$ or $H_{min}$ for larger $a$ to ensure the flatness
of the histogram, but may in the cost of slightly slower convergency, as seen from Fig.~\ref{fig2}.

The new method is extremely general and many techniques developed based on the muticanonical method and/or the WL method could be
applied to the new method.

We thank D. P. Landau, Shan-Ho Tsia, Meng Meng, and Chenggang Zhou for helpful discussion. This work is partly supported by NSFC grant 
No. 11143008.

\begin{table}[h]
\caption{Minimum and maximum path length found by the new method (NM) and the WL method simulations for a TSP that consists of 100 cities.
The average path length and the error bars are calculated from 20 independent runs. See text for more details.}
\label{tab1}
\begin{center}
\begin{tabular}{cccccc}
\hline\hline
 & \multicolumn{2}{c}{Min} & & \multicolumn{2}{c} {Max}\\
MC moves & NM & WL & & NM & WL\\
\hline
$10^6$ & $23.7\pm0.43$ & $23.7\pm0.40$ & & $74.5\pm0.22$ & $74.4\pm0.17$ \\
$10^7$ & $12.7\pm0.23$ & $12.6\pm0.15$ & & $77.38\pm0.08$ & $77.39\pm0.06$ \\
$10^8$ & $8.13\pm0.20$ & $8.13\pm0.06$ & & $77.81\pm0.01$ & $77.81\pm0.01$ \\
\hline
\end{tabular}
\end{center}
\end{table}

\begin{thebibliography}{99}
\bibitem{Ferrenberg} A. M. Ferrenberg and R. H. Swendsen, Phys. Rev. Lett. {\bf 61}, 2635 (1988);
R. W. Swendsen Physica (Amsterdam) {\bf 194A}, 53 (1993).

\bibitem{Berg} B. A. Berg and T. Neuhaus, Phys. Lett. B {\bf 267}, 249 (1991);
Phys. Rev. Lett. {\bf 68}, 9 (1992).

\bibitem{Oliveira} P.M.C. de Oliveira and T.J.P. Penna and H.J. Hermann, Braz. J. Phys., {\bf 26}, 677 (1996);
A. R. Lima, P.M.C. de Oliveira and T.J. P. Penna, J. Stat. Phys., {\bf 99}, 691 (2000).

\bibitem{Lee} J. Lee, Phys. Rev. Lett., {\bf 71}, 211 (1993).

\bibitem{Wang02} F. Wang and D. P. Landau, Phys. Rev. Lett. {\bf 86}, 2050 (2001); Phys. Rev. E {\bf 64}, 056101 (2001); 
Comput. Phys. Commun. {\bf 147}, 674 (2002).

\bibitem{Micheletti} C. Micheletti, A. Laio and  M. Parrinello, Phys. Rev. Lett. {\bf 92}, 170601 (2004); 

\bibitem{Berg96} B. A. Berg, J. Stat. Phys. {\bf 82}, 323 (1996).

\bibitem{Wangswendsen02} J. Wang, and R. H. Swendsen, J. Stat. Phys. {\bf 106}, 245 (2002).

\bibitem{Wust} T. Wust, Comput. Phys. Commun. {\bf 180}, 475 (2009).

\bibitem{Liang} M. Liang, J. Amer. Stat. Asso. {\bf 100}, 472 (2005).

\bibitem{Lin} S. Lin, Bell Syst. Tech. J. {\bf 44}, 2245 (1965).

\bibitem{Belardinelli} R. E. Balardinelli, and V. D. Pereyra, J. Chem. Phys. {\bf 127}, 18 (2007).

\bibitem{Zhou03} C. Zhou and R. N. Bhatt, Phys. Rev. E {\bf 72}, 025701(R) (2005).

\bibitem{Berg97} B. A. Berg,  hep-lat/9705011 (1997).

\end{thebibliography}
\end{document}